\documentclass{iau_FM}
\usepackage{graphicx}

\usepackage{hyperref}

\title[Massive disk galaxies with HSB plus LSB stellar disks ] %% give here short title %%
{Massive disk galaxies with high surface brightness plus low surface brightness stellar disks, hosted by massive dark matter halo \\ - a TNG50 simulation study}

\author[Suchira Sarkar \& Kanak Saha]
{Suchira Sarkar$^{1}$\thanks{email: \texttt{suchira.sarkar@theory.tifr.res.in}}
 \and
 Kanak Saha$^{2}$}

\affiliation
{$^{1}$ Tata Institute of Fundamental Research, Mumbai,
Maharashtra, India\\
$^{2}$ Inter-University Centre for Astronomy and Astrophysics, Pune,
Maharashtra, India}

\pubyear{2024}
\jname{Astronomy in Focus, Volume 1} 
\begin{document}

\maketitle

\begin{abstract}
We study massive disk galaxies (total stellar mass$>=10^{11}$ $\mathrm{M_{\odot}}$) from IllustrisTNG50 simulation, and perform 2-D structural decomposition of the galaxies using their idealised, synthetic SDSS images for z=0. We find an interesting sample of galaxies having a central high surface brightness (HSB) stellar disk, surrounded by an extended low surface brightness (LSB) stellar disk, similar to giant LSB galaxies. These massive, double-exponential disk galaxies are found to be hosted by dark matter haloes of $\sim  10^{12} \mathrm{M_{\odot}}$ in agreement to observations of such galaxies. Their maximum rotation velocity, an approximate measure of their dynamical mass, lies within $\sim$ (300-500) km/s. The stellar-to-dark matter mass ratio and the baryon-to-dark matter mass ratio of the sample lies in the range of $\sim$ (0.04 - 0.46) and $\sim$ (0.07 - 0.47) respectively. Our results show that cosmological simulations are able to form disc galaxies with HSB plus LSB disks, as in observations.
\end{abstract}

\section{Introduction}
The formation of the most massive disk galaxies (total stellar mass $\sim 10^{11} \mathrm{M_{\odot}}$)in the local universe (z$<$0.1) presents an intriguing open question. The giant low surface brightness galaxies are a class of such massive disk galaxies, characterised by a low surface brightness stellar disk (central surface brightness fainter than 22.5 mag.$arcsec^{-2}$ in optical B-band), and a radial exponential scale length of  more than 10 kpc (\cite[Das 2013]{Das2013}). Malin 1, the most extreme example of a gLSB, contains a stellar disk extending upto $\sim$ 100 kpc in radius, and has an extrapolated central surface brightness of $\sim$ 25.5 mag arcsec$^{-2}$ in V- band (\cite[Bothun et al. 1987]{Bothunetal1987}; \cite[Impey \& Bothun 1989]{ImpeyBothun1989}). Interestingly, Malin 1 is shown to have a HSB disk surrounded by a LSB system (\cite[Barth 2007]{Barth2007};\cite[Saha et al. 2021]{Sahaetal2021}). \cite[Saha et al. (2021)]{Sahaetal2021} reports a central HSB, and an outer, extended LSB disk with scale lengths of 4.8 \& 47 kpc, respectively. They report an intermediate LSB disk too. Several other gLSBs with double-exponential disk morphology were discovered (e.g., UGC 1382 (\cite[Hagen et al. 2016]{Hagenetal2016}), NGC 2841 (\cite[Zhang et al. 2018]{Zhangetal2018}), UGC 1378 (\cite[Saburova et al. 2019]{Saburovaetal2019}) etc). \cite[Pandey et al. (2022)]{Pandeyetal2022} found a void galaxy IZw 81, having a smaller radial scale length than a typical giant LSB, showing a similar double-disk morphology. The radial scale length of the exponential HSB and the LSB disks of the above galaxies lie in the range of (2 - 6) kpc, and (7 - 47) kpc, respectively. We aim to search for such massive double-disk galaxies with extended LSB envelope from IllustrisTNG simulation data, and understand their properties (Sarkar \& Saha, 2025). 

\section{Results}
We use the highest resolution version of the cosmological gravo-magnetohydrodynamical simulation suit IllustrisTNG, i,e, TNG50-1, (\cite[Nelson et al. 2019]{Nelsonetal2019}; \cite[Pillepich et al. 2019]{Pillepichetal2019}). The physical simulation box has a cubic volume of roughly 50 Mpc side length. The baryonic and dark matter mass resolution of TNG50-1 are 8.5$\times 10^{4}\mathrm{M_{\odot}}$ \& 4.5$\times 10^{5}\mathrm{M_{\odot}}$, respectively. The gravitational softening for stars and dark matter are 288 parsecs at z = 0, and for gas, the adaptive softening length is minimum 74 comoving parsec. The galaxies are identified as subhaloes while the haloes are identified using a FoF algorithm.
%\subsection{Determination of the double-exponential disk galaxies from IllustrisTNG50}

We find 132 massive galaxies (subhaloes) from TNG50-1 data that have total stellar mass $>=10^{11}\mathrm{M_{\odot}}$. But this sample contains both ellipticals and disk (lenticular and spiral) galaxies. We use idealised, synthetic SDSS-g,r band images of the above sample of galaxies, available in the supplementary data catalogue (\cite[Rodriguez-Gomez et al. 2019]{Rodriguez-Gomezetal2019}) of TNG data to perform all morphological analysis. We filter out the disk galaxies (57) among the above sample, and then perform GALFIT modeling on their images. We find 7 disk galaxies to be best represented by a central sersic plus an inner HSB and an outer, extended LSB exponential disks, based on residual image analysis. We show the results for two such galaxies in Fig. 1. The ranges of the scale length of the HSB disk and LSB disk of the seven galaxies, are obtained to be $\sim$ 2.7-6.0 kpc, and $\sim$ 10.0-30.0 kpc, respectively. We found that, the above structural parameters are similar to that of the observed massive double-disk galaxies as discussed in Section 1.
\begin{figure*}
\centering
\includegraphics[width = 0.34\textwidth, height=0.25\textwidth]{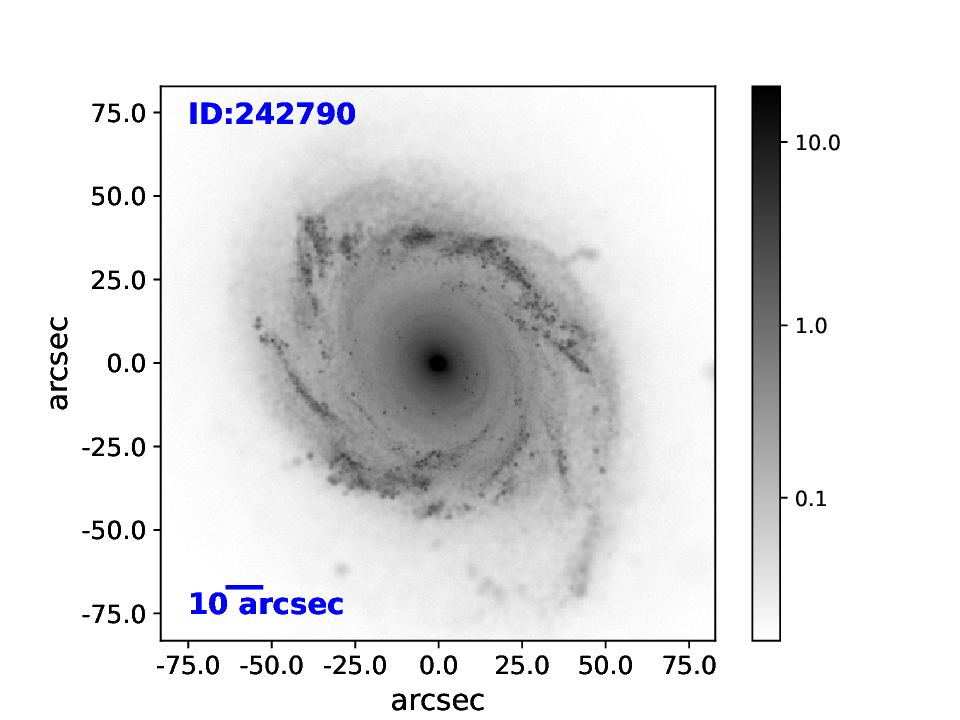}
\includegraphics[width = 0.30\textwidth, height=0.23\textwidth]{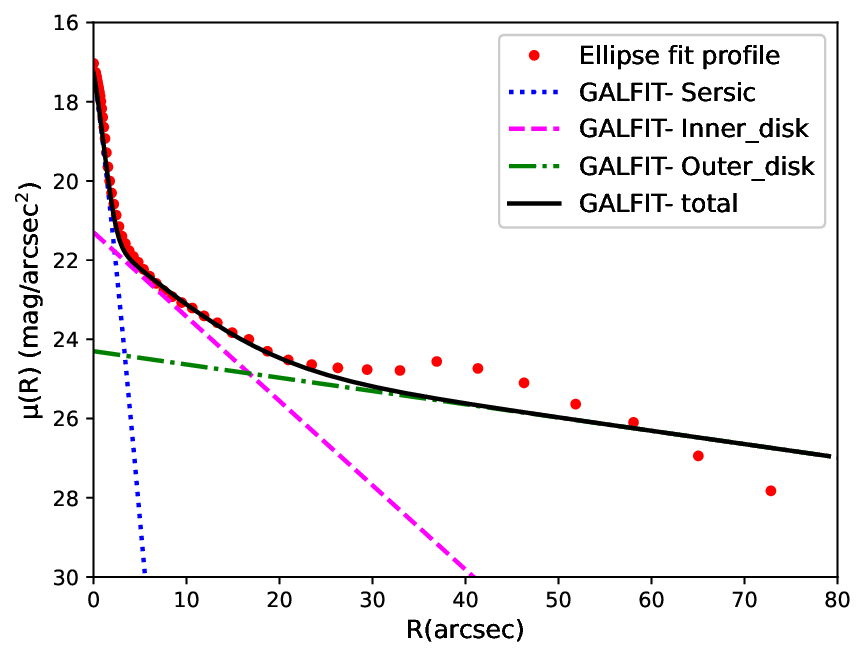}
\includegraphics[width = 0.34\textwidth, height=0.25\textwidth]{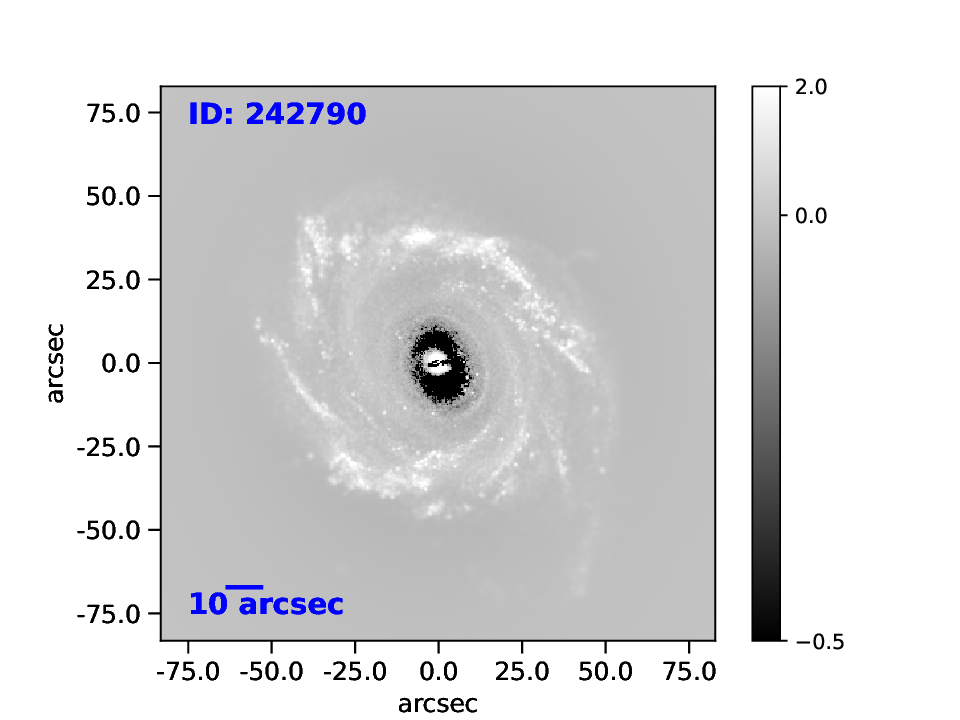}
\medskip
\includegraphics[width = 0.34\textwidth, height=0.25\textwidth]{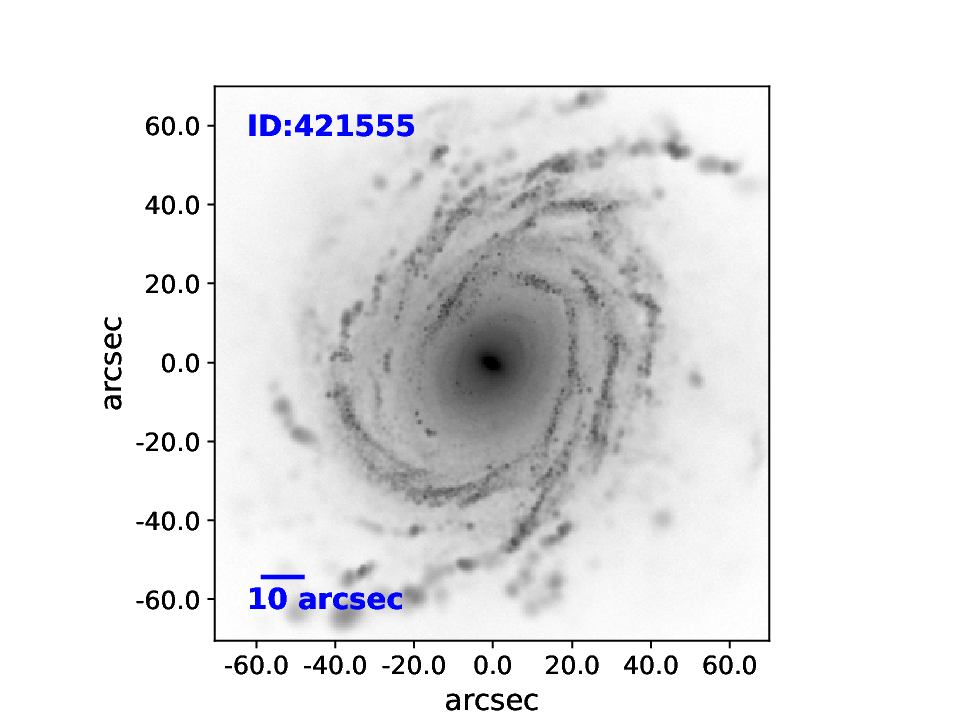}
\includegraphics[width = 0.30\textwidth, height=0.23\textwidth]{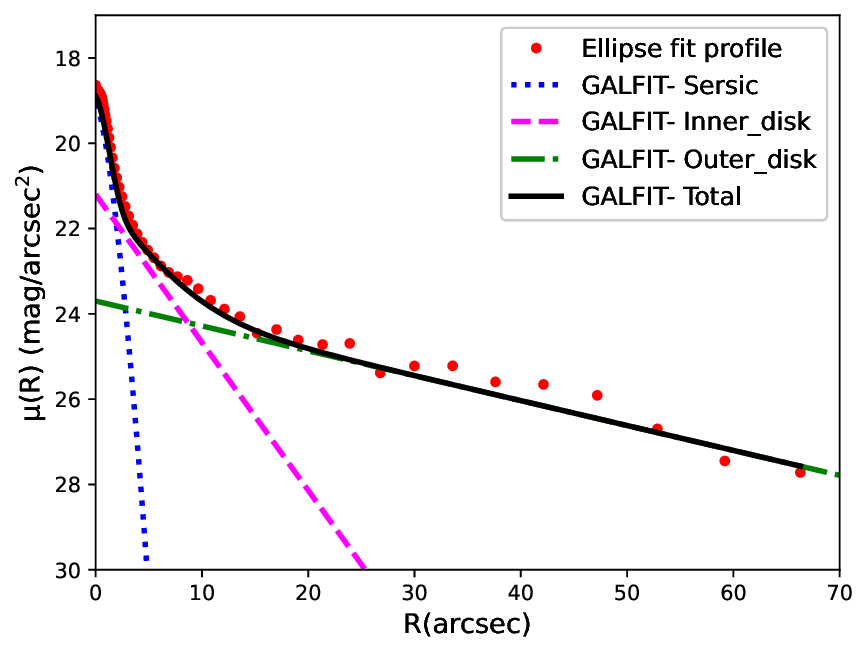}
\includegraphics[width = 0.34\textwidth, height=0.25\textwidth]{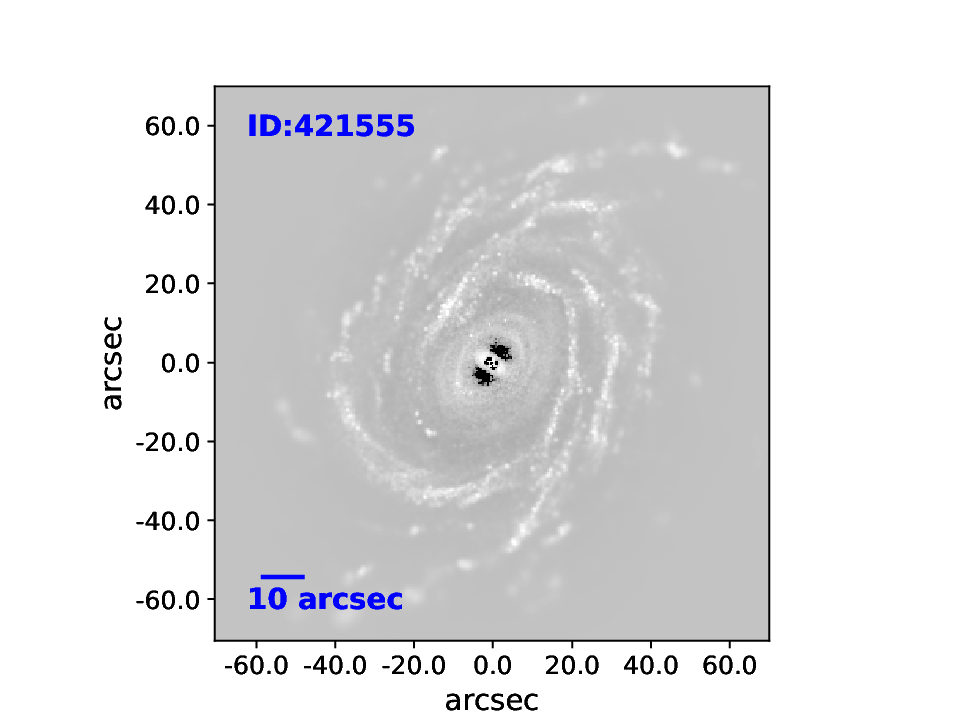}
\caption{Two examples of the massive double-exponential disk galaxies obtained from TNG50-1. \textbf{Left column}: Synthetic SDSS g-band idealised images (displayed in gray scale), shown using the same color bar (in log stretch). Subhalo IDs, i.e., galaxy IDs are mentioned in the images. \textbf{Middle column}: Plot of the 1D surface brightness profiles (red data points) obtained from elliptical isophote fitting of the images. The best-fit central sersic (blue dotted curve), inner (magenta dashed curve) and outer exponential (green dash-dotted curve) disk profiles, calculated analytically from the corresponding best-fit 2D GALFIT model, are over-plotted on the above 1-D surface brightness profile. The total,i.e, the analytical sum of the GALFIT model profiles (black solid curve), agrees well with the 1-D profile of each galaxy. \textbf{Right column}: Residual images obtained from GALFIT modeling, displayed following the same color bar (in log stretch).}
\end{figure*}
%\subsection{Measure of luminous and dark matter mass}

We study the stellar mass, dark matter mass, gas mass, and maximum rotation velocity ($V_{max}$) of the simulated double-disk galaxies, already given in the TNG SUBFIND subhalo data catalogue. The stellar (or dark matter, or gas) mass is calculated as the sum of the mass of all the star (or dark matter, or gas) particles gravitationally bound to the subhaloes. $V_{max}$ is given by the maximum of the spherically averaged rotation curve of the galaxy. The range of stellar mass, dark matter mass, total gas mass are (11.0-11.45), (11.8-12.53), \& (9.87-11.34), respectively, in log($\mathrm{M_{\odot}}$) unit, and the range for $\mathrm{V_{max}}$ is (290-530) $\mathrm{km s^{-1}}$. The stellar-to-dark matter mass ratio is within (0.04-0.46), and baryon(stars+gas)-to-dark matter mass ratio is within (0.07-0.47). We note that, Malin 1 is reported to have a stellar mass of 8.9$\times 10^{11} \mathrm{M_{\odot}}$ (\cite[Saha et al. 2021]{Sahaetal2021}), and HI mass of $6.8 \times 10^{10} \mathrm{M_{\odot}}$ (\cite[Pickering et al. 1997]{Pickeringetal1977}). \cite[Saburova et al. (2019)]{Saburovaetal2019} did mass modeling of UGC 1378 and reported a combined stellar mass of the HSB plus LSB disk as $1.44 \times 10^{11}\mathrm{M_{\odot}}$, a dark matter halo (pseudo-isothermal) mass of $5.52\times 10^{11} \mathrm{M_{\odot}}$ (calculated within 47 kpc). The HI mass and rotation velocity of this galaxy were reported as $1.2\times 10^{10} \mathrm{M_{\odot}}$, and 282 km s$^{-1}$, respectively (\cite[Mishra et al. 2017]{Mishraetal2017}). UGC 1382 (\cite[Hagen et al. 2016]{Hagenetal2016}) is reported to contain a stellar mass (inner lenticular plus outer LSB components) of $8 \times 10^{10} \mathrm{M_{\odot}}$, dark matter halo of mass $2\times10^{12}\mathrm{M_{\odot}}$, HI mass $1.7\times10^{10}\mathrm{M_{\odot}}$, $\mathrm{V_{rot}}=\mathrm{280 km s^{-1}}$. This shows that TNG simulation data is able to reproduce galaxies that have a measure of luminous and dark matter mass approximately comparable to those of similar class of observed galaxies. We note, that for an accurate one-to-one comparison between simulation results and observation, a similar criteria of estimation of mass, Vmax etc. maybe adopted.

\section{Conclusions}
Our results show that the simulations based on $\Lambda$CDM framework can produce massive disk galaxies, with inner high surface brightness plus outer, extended low surface brightness stellar disks, hosted by massive dark matter haloes  ($\sim$ $10^{12}\mathrm{M_{\odot}}$) similar to several observed giant LSB galaxies.

\end{document}